\documentclass[twocolumn,showpacs,preprintnumbers,amsmath,amssymb]{revtex4}
\usepackage{graphicx}% Include figure files

\begin{document}

\title{Ground state magnetic response of two coupled dodecahedra}

\author{N. P. Konstantinidis}
\affiliation{Max Planck Institut f\"ur Physik komplexer Systeme, 01187 Dresden, Germany}

\date{\today}

\begin{abstract}
The antiferromagnetic Heisenberg model on the dodecahedron possesses a number of ground state magnetization discontinuities in a field at the classical and quantum level, even though it lacks magnetic anisotropy. Here the model is considered for two dodecahedra coupled antiferromagnetically along one of their faces, as a first step to determine the magnetic response of collections of fullerene molecules. The magnetic response is determined from the competition among the intra-, interdodecahedral exchange and magnetic field energies. At the classical level the discontinuities of the isolated dodecahedron are renormalized by the interdodecahedral coupling, while new ones show up, with the maximum number of ground state discontinuities being six for a specific range of the coupling. In the full quantum limit where the individual spin magnitude $s=\frac{1}{2}$, there are two ground state discontinuities originating in the single discontinuity of the isolated dodecahedron, and another one due to the intermolecular coupling, generating a total of three discontinuities which come one right after the other. These results show that the magnetic response of more than one dodecahedra interacting together is quite richer than the one of a single dodecahedron.
\end{abstract}


\maketitle

\section{Introduction}
\label{sec:introduction}
The antiferromagnetic (AFM) Heisenberg model (AHM) has been extensively investigated in the recent decades as a prototype for strongly correlated electronic behavior \cite{Auerbach98,Fazekas99}. Special attention has been reserved for lattices and clusters with frustrated connectivity, which in combination with low spatial dimensionality and strong quantum fluctuations can lead to unexpected magnetic behavior \cite{Anderson73,Diep05,Schnack10}. This includes phases without conventional order, such as the spin-liquid phase, non-magnetic excitations inside the singlet-triplet gap, and magnetization plateaux and discontinuities in the response to an external magnetic field.
%Unconventional magnetism is of particular interest for applications such as magnetic storage and quantum computing.

%The magnetism of finite clusters is of particular interest when intermolecular interactions are weak enough, so that experiments directly probe the properties of single molecules. Apart from the practical applications, such molecules are important also from the theoretical point of view, as the validity of models can be thoroughly tested. Such an example is the extensively studied class of single molecule magnets.

In the case of lattices frustration manifests itself in the form of magnetization plateaus  and discontinuities \cite{Honecker04,Nishimoto13,Liu14,Schulenburg02,Nakano13}. Finite clusters can also exhibit magnetization discontinuities. A class of molecules associated with magnetic frustration when the AHM is considered for spins sitting on their vertices are the fullerene molecules. These are hollow carbon molecules that come in the form of closed cages \cite{Fowler95}, with structures that can possess high spatial symmetry. They are made of 12 pentagons and a number of hexagons which varies with the numbers of vertices $n$ as $\frac{n}{2}-10$. Frustration originates in the pentagons and decreases on the average with $n$. The polygons that make up the molecules share their edges, while each vertex is three-fold coordinated. Maybe the most representative member of the class is C$_{60}$, which has the shape of a truncated icosahedron and the spatial symmetry of the largest point symmetry group, the icosahedral group $I_h$. C$_{60}$ was found to superconduct when doped with alkali metals \cite{Hebard91}, and lies in the intermediate $U$ regime of the Hubbard model \cite{Chakravarty91,Stollhoff91}. It was shown that in the large $U$ limit of the Hubbard model, the AHM, the particular connectivity of C$_{60}$ leads to a discontinuity of the magnetization as a function of an external magnetic field in the classical ground state \cite{Coffey92}. This is particularly appealing, as the model lacks any magnetic anisotropy. A classical ground state magnetization discontinuity was also found for the dodecahedron, which is the smallest member of the fullerene class and has 20 vertices and also $I_{h}$ spatial symmetry. The investigation was then extended more generally to molecules of $I_h$ symmetry. First, it was shown that the dodecahedron has in fact a total of three ground state magnetization discontinuities in a field at the classical level, and one and two ground state discontinuities respectively at the full quantum limit of individual spin magnitude $s=\frac{1}{2}$ and 1 \cite{NPK05,NPK07}. It was also shown that the total number of classical ground state magnetization discontinuities for the truncated icosahedron is in fact not one but two, and this is a general feature of fullerene molecules of $I_h$ symmetry. It was established that another general feature of the $I_h$ fullerenes is the high-field discontinuity for $s=\frac{1}{2}$. For relatively small fullerene clusters of different symmetry only pronounced magnetization plateaus were found for $s=1/2$ \cite{NPK09}. In addition the icosahedron, which is not a member of the fullerene family but is the smallest cluster with $I_h$ symmetry, has a classical magnetization discontinuity in its lowest energy configuration which persists for lower values of $s$ \cite{Schroeder05,NPK15}.
%Additionally for small $s$ there are non-magnetic excitations within the singlet-triplet gap and the specific heat has a multi-peak structure as function of temperature for both the dodecahedron and the icosahedron \cite{NPK05}. These are indications of a strong correlation between spatial symmetry and magnetic behavior as the unit that causes frustration is different in the two cases, the pentagon in the fullerenes and the triangle in the icosahedron.
It must also be noted that the on-site repulsion has been found to be stronger for C$_{20}$ than C$_{60}$ in numerical calculations \cite{Lin07,Lin08}, providing further support for the validity of the AHM as a very good approximation of the Hubbard model for the dodecahedron. Work closely related to the above has also been published in the literature \cite{Jimenez14,Hucht11,Strecka15,Sahoo12,Sahoo12-1,Nakano14}.

While C$_{60}$ spontaneously forms in condensation or cluster annealing processes \cite{Kroto87}, this was not the case for C$_{20}$, which was eventually produced in the gas phase by Prinzbach {\it et al.} \cite{Prinzbach00}. C$_{20}$ has been synthesized in the solid phase by Wang {\it et al.}, who produced a hexagonal closed-packed crystal \cite{Wang01}, while Iqbal {\it et al.} synthesized an fcc lattice with C$_{20}$ molecules interconnected by two additional carbon atoms per unit cell in the interstitial tetrahedral sites \cite{Iqbal03}. On the other hand quasi one-dimensional structures have been realized experimentally only a few times for C$_{60}$ molecules, due to their highly anisotropic configuration. This included peapods where C$_{60}$ molecules were introduced in carbon nanotubes, one-dimensional C$_{60}$ structures aligned along step edges on vicinal surfaces of metal single crystals, and chains of C$_{60}$ on self-assembled molecular layers \cite{Smith98,Zhang07,Tamai04,Zeng01}. Most recently C$_{60}$ molecules were arranged on chain structures of width two to three molecules on rippled graphene \cite{Chen15}. In addition, as few as two C$_{60}$ molecules have been considered to link and form a dumbbell structure \cite{Manaa09}.

The formation of $I_h$-fullerene lattice structures poses the question of the influence of intermolecular interactions on isolated molecule properties. Considering interactions again at the level of the AHM, it is of interest to determine if the appealing ground state magnetic features of a single dodecahedron survive in a lattice-type setting, and if the addition of intermolecular magnetic exchange introduces extra features in the magnetization. This is the main question undertaken in this paper. More specifically the case of two dodecahedra is investigated, with intramolecular interactions exactly as in the isolated dodecahedron case, while the two dodecahedra are connected along with one of their faces with a varying exchange interaction. The properties of the ground state of the whole cluster are mapped as a function of the intermolecular exchange constant and an external magnetic field. For weak intermolecular coupling the response is mainly determined by the isolated dodecahedra, while for strong by the dimer-type interaction between spins belonging to different dodecahedra.

For relatively small intermolecular coupling the three classical discontinuities of the lowest energy configuration of the isolated dodecahedron survive. Simultaneously a new low-field magnetization discontinuity appears, which relates to the AFM coupling of the two molecules and its competition with the magnetic field. This discontinuity survives up to the dimer limit. For stronger coupling even more discontinuities appear, producing a rich structure for the classical magnetic response of the two dodecahedra system. For a specific range of interdodecahedral coupling the total number of ground state discontinuities goes up to six. The spins associated with the interdodecahedral interaction do not necessarily increase their projection along the field as the latter increases, due to their unfrustrated AFM interaction. Finally, one of the discontinuities becomes one of the susceptibility close to the dimer limit.

In the $s=\frac{1}{2}$ case the isolated dodecahedron discontinuity generates two ground state discontinuities for the coupled dodecahedra. In addition, a third discontinuity appears due to the interdodecahedral coupling. All three discontinuities persist for smaller values of the interdodecahedral coupling and appear one right after the other. This generates a ground state magnetic response which for a considerable range of magnetic fields is associated with magnetization discontinuities of the total spin along the $z$ axis equal to $\Delta S^z=2$, instead of the typical $\Delta S^z=1$ for quantum mechanics.

The plan of this paper is as follows: In Sec. \ref{sec:model} the AHM for the system of the two dodecahedra is introduced, and in Sec. \ref{sec:classicalspins} its lowest energy configuration is calculated for classical spins. Section \ref{sec:spinsonehalf} considers the $s=\frac{1}{2}$ case, with perturbation theory for weak and Lanczos diagonalization for arbitrary values of the interdodecahedral coupling. Finally Sec. \ref{sec:conclusions} presents the conclusions.

\section{Model}
\label{sec:model}
\begin{figure}
\includegraphics[width=3.5in,height=2.6in]{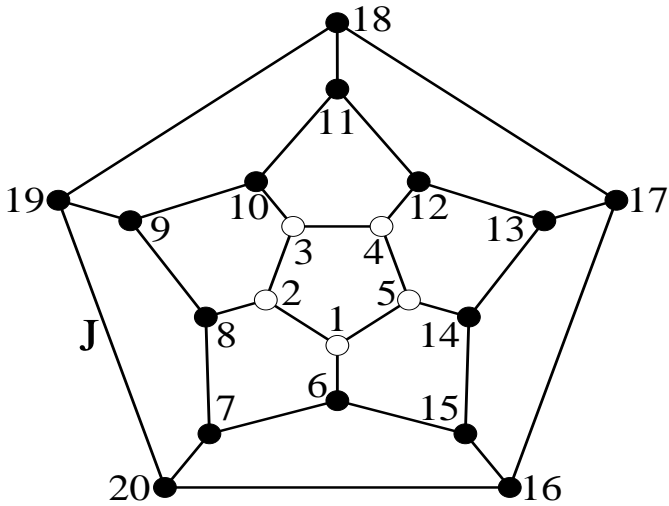}
\caption{Projection of a single dodecahedron on a plane. The solid lines are antiferromagnetic bonds with strength $J$ and the circles spins with magnitude $s$. The cluster considered in this paper consists of two such dodecahedra connected with $J'$ interactions between the spins of one of their faces, highlighted with white color and including spins 1 to 5. The spins of the second dodecahedron have the same linking pattern with the first as shown in the figure and their indices are given as $20+i$, where $i$ runs from 1 to 20. The $J'$ bonds are between spins $i$ and $20+i$, with $i$ running from 1 to 5.
%(/basic/LATEX)
}
\label{fig:twododecahedra}
\end{figure}

The AHM for two linked dodecahedra (Fig. \ref{fig:twododecahedra}) is:
\begin{eqnarray}
H & = & J ( \sum_{<ij>} \textrm{} \vec{s}_i \cdot \vec{s}_j + \sum_{<20+i,20+j>} \textrm{} \vec{s}_{20+i} \cdot \vec{s}_{20+j} ) \nonumber \\ & & + J' \sum_{i=1}^5 \vec{s}_i \cdot \vec{s}_{20+i} - h \sum_{i=1}^{N} s_i^z
\label{eqn:Hamiltonian}
\end{eqnarray}
The total number of spins for the two dodecahedra is $N=40$, with the first dodecahedron containing spins 1 to 20 and the second 21 to 40. The first two sums run over nearest-neighbor spins within the same dodecahedron, with $i$ and $j$ running from 1 to 20. The second sum connects the spins on the faces of the two dodecahedra that are taken to be directly opposite each other, with $i$ an index counting the spins on these two faces. The magnetic field $h$ is taken to be directed along the $z$ axis. The system interpolates between two independent dodecahedra for $J'=0$, and five independent dimers for $J=0$. The ratio $\alpha \equiv \frac{J'}{J+J'}$ is defined, which correspondingly varies between 0 and 1.

\section{Classical Spins}
\label{sec:classicalspins}
First the spins of Hamiltonian (\ref{eqn:Hamiltonian}) are taken to be classical \cite{NPK15-1}. When $J'=0$ the ground state magnetization of each isolated dodecahedron has three discontinuities in the field, occuring when $\frac{h}{h_{sat}^{dod}}=0.26350$, 0.26983, and 0.73428, with $h_{sat}^{dod}=(3+\sqrt{5})J$ the saturation field of an isolated dodecahedron \cite{NPK07}. The symmetry of the ground state configuration does not necessarily increase with the magnetic field. In zero field nearest-neighbor spins are not antiparallel due to frustration, and the nearest-neighbor correlation in each dodecahedron equals $-\frac{\sqrt{5}}{3}$. On the other hand, if the $J'$ bonds between different dodecahedra were to be considered alone their spins would be antiparallel in the lowest energy state, consequently the non-frustrated dimer bonds should be less susceptible to an external field in comparison with the frustrated intradodecahedral bonds.

\begin{figure}
\includegraphics[width=3.5in,height=2.6in]{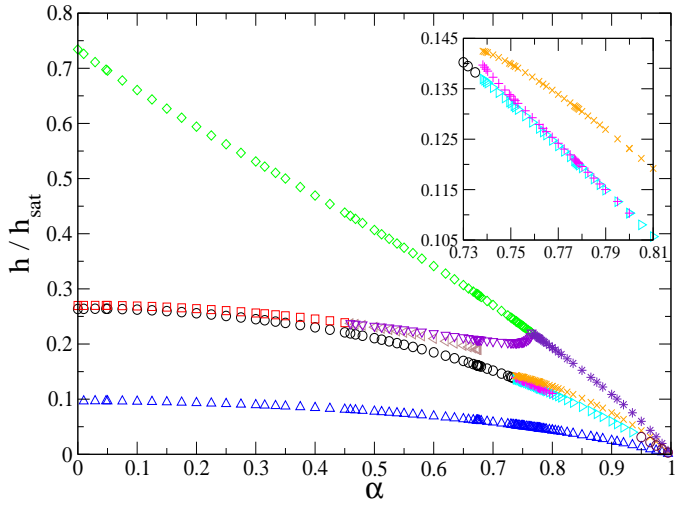}
\caption{(Color online) Location of the classical ground state discontinuities with respect to $\alpha$ and $\frac{h}{h_{sat}}$. Each discontinuity is distinguished by a different symbol (and color). The (black) circles, (red) squares and (green) diamonds correspond to the magnetization discontinuities that originate in the isolated dodecahedron ($J'=0$). The (maroon) circles on the far right correspond to the sole susceptibility discontinuity. The inset shows in detail the three magnetization discontinuities that are close to each other.
%(basic/classical/dodecahedronarray/two)
}
\label{fig:configurationdiagram}
\end{figure}

\begin{figure}
\includegraphics[width=2.6in,height=3.5in,angle=270]{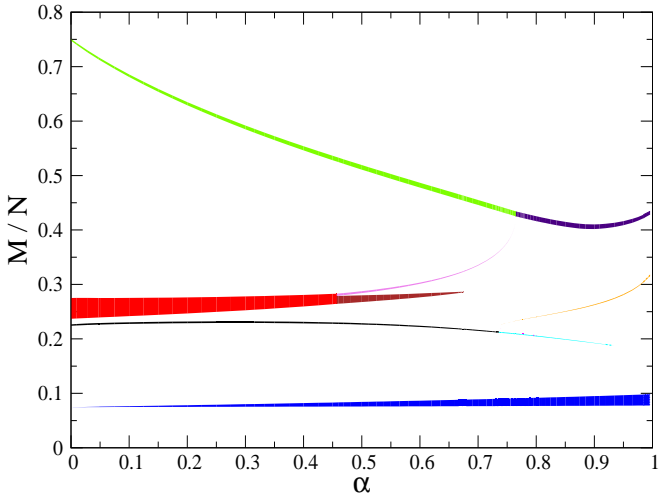}
\caption{(Color online) Inaccessible lowest energy state magnetizations per spin $\frac{M}{N}$ as functions of $\alpha$. The inaccessible magnetizations are distinguished by their color, in accordance with Fig. \ref{fig:configurationdiagram}.
%(basic/classical/dodecahedronarray/two)
}
\label{fig:discontinuitiesminmax-a}
\end{figure}

\begin{figure}
\includegraphics[width=3.5in,height=2.6in]{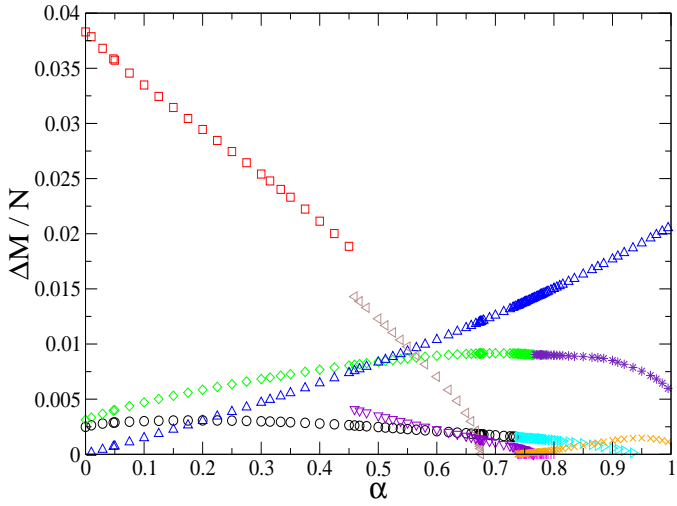}
\caption{(Color online) Magnetization change per spin $\frac{\Delta M}{N}$ for the classical ground state magnetization discontinuities as a function of $\alpha$. Each discontinuity is distinguished by a different symbol (and color), in accordance with Figs. \ref{fig:configurationdiagram} and \ref{fig:discontinuitiesminmax-a}.
%(basic/classical/dodecahedronarray/two)
}
\label{fig:discontinuitieswidth}
\end{figure}

For finite $J'$ and zero field the relative spin orientations in each dodecahedron in the ground state do not change with respect to the noninteracting case, while spins connected via $J'$ bonds align themselves in an antiparallel fashion and the energy is $-20 \sqrt{5}J - 5J'$. Once the magnetic field is switched on the competition among the intra-, interdodecahedral exchange and magnetic field energies determines the lowest energy configuration. The magnetization discontinuities associated with an isolated dodecahedron survive the interdodecahedral coupling, while new ones emerge. The location of all the lowest energy configuration discontinuities with respect to $\alpha$ and $\frac{h}{h_{sat}}$ is shown in Fig. \ref{fig:configurationdiagram} ($h_{sat}$ is the saturation field of the two dodecahedra, which is a function of $\frac{J'}{J}$). Apart from $\alpha$ close to 1, where the second discontinuity with respect to the field strength becomes the sole susceptibility discontinuity, there are never less than four magnetization discontinuities, showing that the introduction of the interaction between the dodecahedra enriches the magnetic response for any coupling strength. The maximum number of magnetization discontinuities occurs for $\alpha \sim \frac{3}{4}$ and is equal to six. The discontinuities occur mostly for decreasing $\frac{h}{h_{sat}}$ with increasing $\alpha$, until they eventually disappear in the dimer limit $\alpha=1$. The inaccessible magnetizations per spin which fall between the edges of each discontinuity are plotted in Fig. \ref{fig:discontinuitiesminmax-a}. The corresponding magnitudes of the magnetization change per spin are shown in Fig. \ref{fig:discontinuitieswidth}. The width of the nonaccessible magnetizations is not necessarily monotonic with $\alpha$.

\begin{figure}
\includegraphics[width=3.5in,height=2.6in]{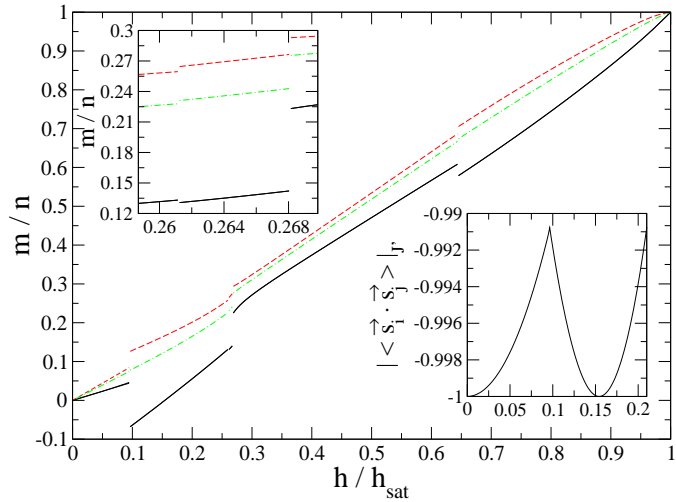}
\vspace{0pt}
\caption{(Color online) Total classical ground state magnetization projection along the $z$ axis per spin $\frac{m}{n}$ as a function of $\frac{h}{h_{sat}}$ for $\frac{J'}{J}=\frac{1}{7}$ ($\alpha=\frac{1}{8}$). The (black) continuous line corresponds to the spins in each dodecahedron connected via $J'$ bonds, the (red) long-dashed line to the rest of the spins, and the (green) long-dashed-dot line to all the spins. The lower right inset shows the average interdodecahedral correlation function $|<\vec{s}_i \cdot \vec{s}_j>|_{J'}$ for small fields. The upper left inset shows in detail two of the discontinuities.
%(basic/classical/dodecahedronarray/two/J1=1J2=1over7/run5)
%(basic/classical/dodecahedronarray/two/J1=1J2=1over7/run10)
}
\label{fig:magnetizationpartsinterexchangeJ1=1J2=1over7}
\end{figure}

Once $J'$ becomes non-zero, apart from the three discontinuities of $J'=0$ a new one appears for small magnetic fields, where the two dodecahedra are still connected approximately in an antiparallel fashion via the $J'$ bonds. This discontinuity increases its strength monotonically with $J'$, reaching $\Delta M \sim 0.8$ close to the dimer limit (Figs. \ref{fig:discontinuitiesminmax-a} and \ref{fig:discontinuitieswidth}). The magnetization curve for $\frac{J'}{J}=\frac{1}{7}$ ($\alpha=\frac{1}{8}$) is shown in detail in Fig. \ref{fig:magnetizationpartsinterexchangeJ1=1J2=1over7}. For small fields the spins associated with the non-frustrated bonds on the average do not respond as strongly as the rest, maintaining a smaller total projection along the $z$ axis and having interdodecahedral correlations only weakly deviating from -1 (lower right inset of Fig. \ref{fig:magnetizationpartsinterexchangeJ1=1J2=1over7}). The rest of the nearest-neighbor correlations deviate more strongly from their zero field value, and this deviation increases with $h$. For $\frac{h}{h_{sat}}=0.095350$ the magnetization discontinuity originating in the interdodecahedral coupling appears. The lowest energy configuration right after the discontinuity is similar to the lower-field ground state configuration of an isolated dodecahedron \cite{NPKLuban}.
%The new configuration is more symmetric as the spins belong to one of four distinct polar angle values, while the azimuthal angles are equidistantly distributed by $\frac{\pi}{5}$ in groups of four spins.
Even though the total magnetization of the two dodecahedra increases right after the low-field discontinuity, the net magnetization of the spins connected via $J'$ bonds changes direction and points away from the field. These spins now share a common polar angle which starts out bigger than $\frac{\pi}{2}$, and then monotonically decreases with the field. The five $J'$ correlations are equal and become more antiferromagnetic with increasing field (lower right inset of Fig. \ref{fig:magnetizationpartsinterexchangeJ1=1J2=1over7}). For a specific value of the field the common polar angle becomes equal to $\frac{\pi}{2}$, and then the $J'$ bonds connect antiparallel spins which have zero net magnetization. If the field is further increased the polar angle of the $J'$ spins becomes less than $\frac{\pi}{2}$ and they start to deviate from being antiparallel, while their net magnetization is now non-zero and points towards the field.

Apart from the low-field discontinuity, the net magnetization of the $J'$ spins decreases also at the second and last ones (Fig. \ref{fig:magnetizationpartsinterexchangeJ1=1J2=1over7} and its upper left inset). The three higher-field discontinuities in Fig. \ref{fig:magnetizationpartsinterexchangeJ1=1J2=1over7} are directly related to the ones of an isolated dodecahedron \cite{NPK07}, only renormalized by the interdodecahedral interaction. One of them, the middle one, is the strongest among all discontinuities, and for small $J'$ it is associated with a magnetization change $\Delta M \sim 1.5$ (Figs. \ref{fig:discontinuitiesminmax-a} and \ref{fig:discontinuitieswidth}).

\begin{figure}
\includegraphics[width=3.5in,height=2.6in]{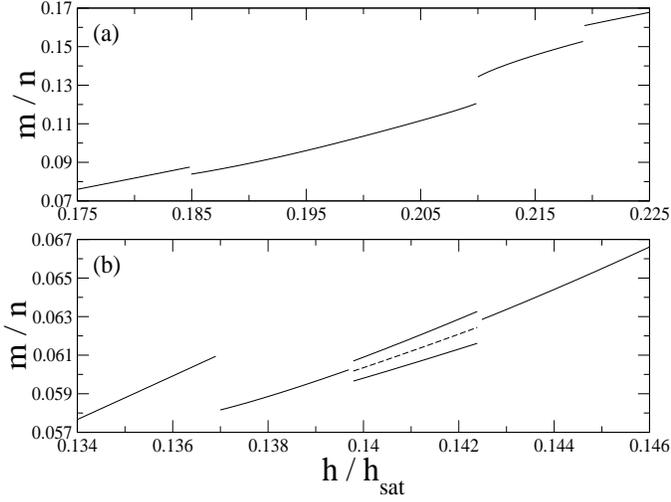}
\vspace{0pt}
\caption{Total classical ground state magnetization projection along the $z$ axis per spin $\frac{m}{n}$ as a function of $\frac{h}{h_{sat}}$ for the spins in each dodecahedron connected via $J'$ bonds for (a) $\frac{J'}{J}=\frac{3}{2}$ ($\alpha=\frac{3}{5}$), and (b) $\frac{J'}{J}=2.82$ ($\alpha=0.73822$). The long-dashed line gives the average when the two values differ.
%(basic/classical/dodecahedronarray/two)
%(basic/classical/dodecahedronarray/two/J1=1J2=3over2/run2)
%(basic/classical/dodecahedronarray/two/J1=1J2=2.82/run)
}
\label{fig:magnetizationpartsinterexchangeJ1=1J2=3over2J1=1J2=2.82}
\end{figure}

When $\frac{J'}{J}=0.84139$ ($\alpha=0.45693$) the third discontinuity splits up in two (Fig. \ref{fig:configurationdiagram}), which have smaller magnitude (Figs. \ref{fig:discontinuitiesminmax-a} and \ref{fig:discontinuitieswidth}). The net magnetization of the $J'$ spins is shown in Fig. \ref{fig:magnetizationpartsinterexchangeJ1=1J2=3over2J1=1J2=2.82}(a) for $\frac{J'}{J}=\frac{3}{2}$ ($\alpha=\frac{3}{5}$). The two discontinuities result in a stepwise increase of the magnetization of the $J'$ spins, as seen in the right part of the figure. The rest of the spins also increase their net magnetization stepwise. The lower of these two discontinuities disappears for $\frac{J'}{J}=2.0800$ ($\alpha=0.67532$) (Fig. \ref{fig:configurationdiagram}). For $\frac{J'}{J}=2.78174$ ($\alpha=0.73557$) and right above the second discontinuity two new discontinuities emerge (Fig. \ref{fig:configurationdiagram} and inset), bringing the total number to a maximum equal to six. The net magnetization of the $J'$ spins is plotted in Fig. \ref{fig:magnetizationpartsinterexchangeJ1=1J2=3over2J1=1J2=2.82}(b) for the case $\frac{J'}{J}=2.82$ ($\alpha=0.73822$). Here the net magnetization of the right- and left-dodecahedron $J'$ spins is not the same after the first new discontinuity, as seen in the right part of the figure, and the net magnetization of the $J'$ spins decreases when this difference appears. The top two discontinuities merge for $\frac{J'}{J}=3.2552$ ($\alpha=0.76499$) (Fig. \ref{fig:configurationdiagram}), while for $\frac{J'}{J}=3.99890$ ($\alpha=0.79996$) the second and the third discontinuity merge. Finally for $\frac{J'}{J} \sim 17.1$ ($\alpha \sim 0.945$) the second discontinuity changes from a magnetization to a susceptibility discontinuity.

The lowest energy configuration spin directions change discontinuously like the magnetization as the gaps are encountered with increasing field. Below the first magnetization discontinuity the spin configuration is highly asymmetric, with each spin having its own polar angle. Low symmetry is a general feature of the configurations of Fig. \ref{fig:configurationdiagram}, as the spins at best have their own polar angle value within an individual dodecahedron, with each of these polar angle values shared only by another spin in the other dodecahedron. The most symmetric of these configurations is when the spins mounted exactly at the same location in the two dodecahedra share the polar angle, and their azimuthal angles differ by $\pi$. The most notable exception to these cases is the lowest energy configuration right after the low-field magnetization jump, which is indicated with (blue) up triangles in Fig. \ref{fig:configurationdiagram}, which is also the last configuration just before saturation, occuring for fields higher than the ones depicted with (green) diamonds. This configuration is shown in Fig. \ref{fig:twododecahedraquantum}.
%paperseven-1new.eps.
Similarly to the low-field ground state configuration of an isolated dodecahedron \cite{NPKLuban}, there are four distinct polar angles for the spins, with each one corresponding to a different circle type. Lines of the same type represent equal nearest-neighbor correlations. All the azimuthal angles are integer multiples of $\frac{\pi}{5}$, while successive azimuthal angles within the same theta group differ by $\frac{4\pi}{5}$. Along the central line defined by spins 1, 6, 11, and 18, nearest-neighbors differ by $\pi$ in the azimuthal plane. Spins symmetrically placed with respect to this line have azimuthal angles adding up to $2\pi$. The polar angles are the same for spins placed exactly at the same locations in the two dodecahedra, while their azimuthal angles differ by $\pi$. In the lowest energy configuration before the (orange) x's in Fig. \ref{fig:configurationdiagram} the polar angles are different in the two dodecahedra, and there are 12 distinct polar angles for each one of them.

\section{$s=\frac{1}{2}$}
\label{sec:spinsonehalf}
For $s=\frac{1}{2}$ an isolated dodecahedron has a discontinuity in the ground state magnetization where its total $z$ spin sector $S_{dod}^z=5$ with five spin flips from saturation is never the ground state in a field \cite{NPK05}. This results from the energy difference of the $S_{dod}^z=5$ and 6 lowest energy states being smaller than the one of the $S_{dod}^z=4$ and 5 lowest energy states. The discontinuity carries over to the case of the two linked dodecahedra: their lowest energy wavefunction for a specific $S^z$ when $J'=0$ is the product of the individual dodecahedra ground state wavefunctions for specific $S_{dod}^z$'s that minimize the energy and have spins adding up to $S^z$, with the total energy equaling the sum of the corresponding individual dodecahedra energies. As a result the isolated dodecahedron discontinuity generates two discontinuities for the coupled dodecahedra, where the lowest energy levels with $S^z=9$ and $S^z=11$ are never the ground states in a field. When $J'$ becomes non-zero, the discontinuities will survive at least for weak values of it. The ground state magnetization curve of Hamiltonian (\ref{eqn:Hamiltonian}) is calculated for the whole $S^z$ range when $J'$ is weak with perturbation theory. The magnetization is also calculated with Lanczos diagonalization for arbitrary $J'$, however due to computational requirements it can not be determined for lower $S^z$ in this case.

\subsection{Perturbation Theory}
\label{subsec:perturbationtheory}

For small $J'$ the lowest energies are calculated for every $S^z$ sector within first order perturbation theory. The unperturbed wavefunction ($J'=0)$ is the product of the lowest energy wavefunctions of the two dodecahedra according to the $S_{dod}^z$ sector they belong to (see Table V of Ref. \cite{NPK05}). When $S_{dod}^z$ is away from the single dodecahedron discontinuity for both dodecahedra, the zeroth order wavefunction for even $S^z$ is the product of the lowest energy state with $S_{dod}^z=\frac{S^z}{2}$ for each dodecahedron:
\begin{eqnarray}
\vert \Psi_0^{i*d+j}(S^z) \rangle_{J'=0} = \vert \Phi_0^i(\frac{S^z}{2}) \rangle \vert \Phi_0^j(\frac{S^z}{2}) \rangle
\label{eqn:pertheoryevenwavefunction}
\end{eqnarray}
The index $i$, $j=1,\dots,d$ counts the degeneracy $d$ of the single dodecahedron wavefunction, therefore the unperturbed wavefunction of the two dodecahedra is in principle also degenerate. For odd $S^z$ the combining single dodecahedron lowest energy states have $S_{dod}^z=\frac{S^z-1}{2}$ and $\frac{S^z+1}{2}$. Here apart from the degeneracy originating in the degeneracy of the single dodecahedron lowest energy states, an extra factor of two comes about from the two distinct $S_{dod}^z$ values to be accomodated on the two dodecahedra. The only exception in the general pattern is $S^z=10$, where the participating single dodecahedron lowest energy states do not have the same $S_{dod}^z=5$ as in the other even cases, but due to the single dodecahedron discontinuity they have $S_{dod}^z=4$ and 6. Thus in general degenerate perturbation theory is required, unless $S^z$ is even and different from 10 and the lowest single dodecahedron energy level for $\frac{S^z}{2}$ is singly degenerate. The perturbative term is scaled with $J'$ in Hamiltonian (\ref{eqn:Hamiltonian}), and first order degenerate perturbation theory produces the following matrix, exemplarily for the even $S^z$ case of Eq. (\ref{eqn:pertheoryevenwavefunction}):
\begin{eqnarray}
& & H_1 (i*d+j,k*d+l) = \nonumber \\ & & \textrm{ }_{J'=0}\langle \Psi_0^{i*d+j}(S^z) \vert \sum_{m=1}^5 \vec{s}_m \cdot \vec{s}_{20+m} \vert \Psi_0^{k*d+l}(S^z) \rangle_{J'=0} \nonumber \\
\end{eqnarray}
The perturbative term includes combinations of raising and lowering operators, as well as diagonal terms. For even $S^z \neq 10$ only diagonal terms can generate non-zero energy contributions in first order perturbation theory, irrespective of the degeneracy of the single dodecahedron state $\vert \Phi_0^i(\frac{S^z}{2}) \rangle$. For odd $S^z$ and $S^z=10$ combinations of raising and lowering operators also contribute.

The first order perturbation theory correction for the energy $E_1$ is listed for the different $S^z$ sectors in Table \ref{table:firstordercorrection}. According to what has already been mentioned in this Subsection about the $S_{dod}^z$ sectors that combine to form the unperturbed wavefunction, away from the discontinuity the energy difference between levels $S^z$, $S^z-1$ and $S^z-1$, $S^z-2$ is the same for $J'=0$ when $S^z$ is even. Consequently what determines the relative value of successive energy differences between adjacent $S^z$ sectors for weak $J'$ are the perturbative energy corrections listed in Table  \ref{table:firstordercorrection}. If the relative energy difference between three successive $S^z$ sectors in Table \ref{table:firstordercorrection} when starting from an even $S^z$ increases for decreasing $S^z$, then a new magnetization discontinuity appears. This is the case for $S^z=14$, and the magnetization in a field switches from $S^z=14$ directly to $S^z=12$ at least for small $J'$. This discontinuity does not relate to the one of the isolated dodecahedron but originates in the interdodecahedral coupling. It is then concluded that at least for weak $J'$ and between $S^z=8$ and 14 the magnetization changes with an external field in steps of $\Delta S^z=2$. This shows that for two linked dodecahedra the magnetization can be changed in a controlled way in steps of either $\Delta S^z=1$ or 2 by adjusting the range of an external magnetic field.

\subsection{Lanczos Diagonalization}
\label{subsec:lanczosdiagonalization}

The magnetization response of Hamiltonian (\ref{eqn:Hamiltonian}) can be calculated for $J'$ of arbitrary strength with Lanczos diagonalization, taking into account the $D_{5h}$ spatial symmetry of the Hamiltonian \cite{NPK05,NPK07,NPK09,NPK15,NPK04}. In this way the Hamiltonian is block-diagonalized according to the irreducible representations of its symmetry group. This results in eigenstates well-defined according to symmetry, as well as a Hamiltonian divided in smaller subblocks that are easier to diagonalize. In contrast with the perturbation theory calculation of Sec. \ref{subsec:perturbationtheory} here there is no restriction on the strength of $J'$, however the calculation is limited to higher values of $S^z$ due to computational requirements.

\begin{figure}
\includegraphics[width=3.5in,height=2.6in]{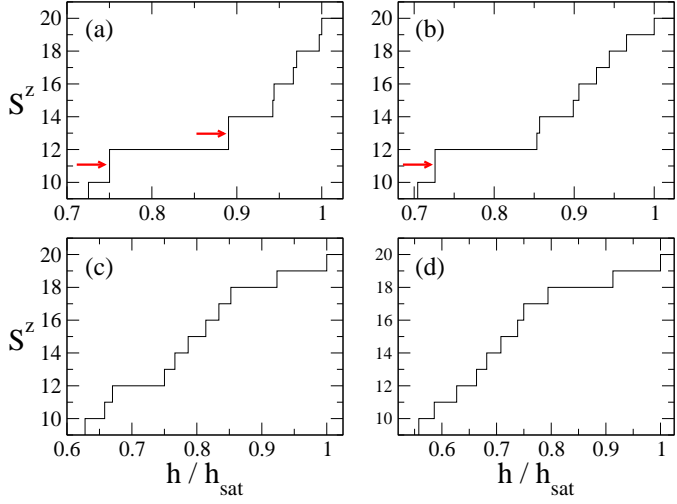}
\vspace{0pt}
\caption{(Color online) $S^z$ in the ground state as a function of $\frac{h}{h_{sat}}$ for $s=\frac{1}{2}$ and $J'$ equal to (a) $\frac{J}{50}$, (b) $\frac{J}{5}$, (c) $\frac{3J}{5}$, and (d) $J$. The (red) arrows point at the magnetization discontinuities, where $\Delta S^z=2$.
%(/basic/diag/dodectwo)
}
\label{fig:magnetizationquantum}
\end{figure}

%$J'=0.02$, 0.2, 0.6, 1.
Fig. \ref{fig:magnetizationquantum} shows the ground state magnetization curve for four different $J'$ values. When $J'$ is small three successive discontinuities are expected according to Sec. \ref{subsec:perturbationtheory}, with the subsectors $S^z=9$, 11 and 13 never including the lowest energy state in a magnetic field. Two discontinuities are highlighted with (red) arrows in Fig. \ref{fig:magnetizationquantum}(a) where $J'=\frac{J}{50}$, while the one that corresponds to the lowest $S^z$ is not mapped out due to the computational requirements to find the lowest energy state for $S^z=8$. When $J'=\frac{J}{5}$ (Fig. \ref{fig:magnetizationquantum}(b)) the interdodecahedral coupling is not weak any more, and the highest $S^z$ discontinuity has disappeared. For even higher $J'=\frac{3J}{5}$ (Fig. \ref{fig:magnetizationquantum}(c)) the dodecahedra feel each other's influence more strongly, and as a result the intermediate discontinuity disappears as well. Fig. \ref{fig:magnetizationquantum}(d) shows the magnetization for $J'=J$ where all couplings are equal and even magnetization plateaus are absent for the $S^z$ range of the weaker $J'$ discontinuities.

\begin{figure}
\includegraphics[width=3.5in,height=2.6in]{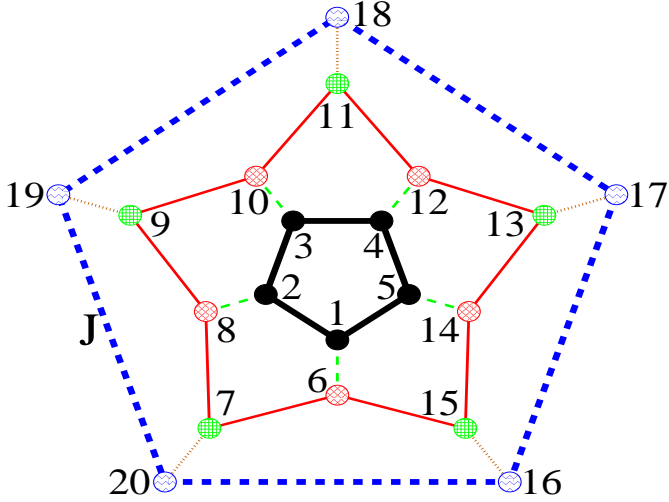}
\caption{(Color online) Projection of a single dodecahedron on a plane (see Fig. \ref{fig:twododecahedra} for description). Lines of the same type (and color) represent equal nearest-neighbor correlations $\vec{s}_i \cdot \vec{s}_j$ in the lowest-energy configuration right after the low-field magnetization discontinuity and just before saturation for classical spins, and equal nearest-neighbor correlations $<\vec{s}_i \cdot \vec{s}_j>$ in the ground state for $s=\frac{1}{2}$ for different $S^z$. Circles of the same pattern (and color) represent equal polar angles in the lowest-energy configuration right after the low-field magnetization discontinuity and just before saturation for classical spins, and equal projections of local spins along the $z$ axis $<s_i^z>$ in the ground state for $s=\frac{1}{2}$ for different $S^z$.
%(/basic/LATEX)
}
\label{fig:twododecahedraquantum}
\end{figure}

\begin{figure}
\includegraphics[width=3.5in,height=2.6in]{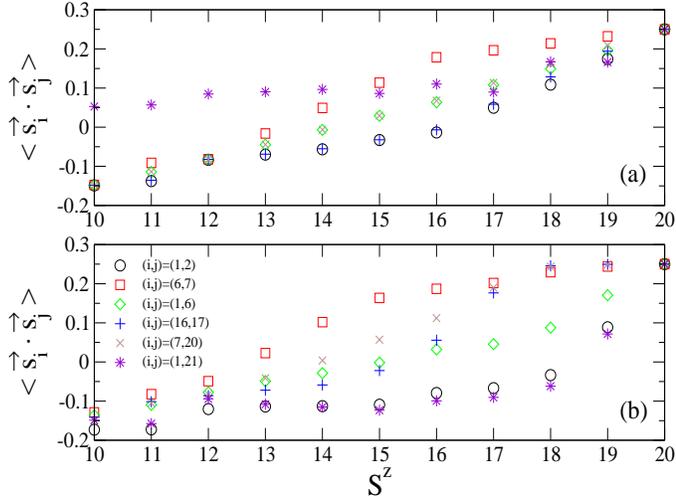}
\vspace{0pt}
\caption{(Color online) Distinct ground state expectation values of the nearest-neighbor correlation functions $<\vec{s}_i \cdot \vec{s}_j>$ as a function of $S^z$ for $s=\frac{1}{2}$ and $J'$ equal to (a) $\frac{J}{50}$, and (b) $\frac{3J}{5}$. The spin indices of the legend (and the color coding) refer to Fig. \ref{fig:twododecahedraquantum}.
%(/basic/diag/dodectwo)
}
\label{fig:correlationsquantum}
\end{figure}

Fig. \ref{fig:correlationsquantum} shows the distinct ground state expectation values of the nearest-neighbor correlation functions $<\vec{s}_i \cdot \vec{s}_j>$ for $J'=\frac{J}{50}$ and $\frac{3J}{5}$. There are in principle six unique such correlations, with one for each of the rings that respectively contain spins 1 to 5, 6 to 15, and 16 to 20 (Fig. \ref{fig:twododecahedraquantum}), and two more for nearest-neighbor correlations between spins that belong to different rings. Tha last unique correlation is between spins belonging to different dodecahedra. For $J'=\frac{J}{50}$ the single-dodecahedron character is preserved, at least not for high $S^z$, where only correlations within the two dodecahedra are antiferromagnetic. The intradodecahedral correlation (represented by $<\vec{s}_1 \cdot \vec{s}_{21}>$ in Fig. \ref{fig:correlationsquantum}) is ferromagnetic for all the $S^z$ presented. The situation is different when $J'=\frac{3J}{5}$. Now the intradodecahedral correlation acquires an AFM character, which is the strongest along with its neighboring correlation $<\vec{s}_1 \cdot \vec{s}_2>$.

\begin{figure}
\includegraphics[width=3.5in,height=2.6in]{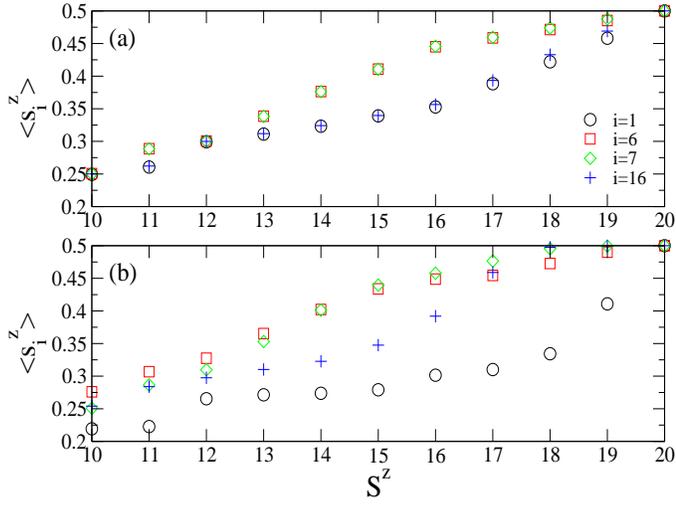}
\vspace{0pt}
\caption{(Color online) Distinct ground state expectation values of the projections of local spins along the $z$ axis $<s_i^z>$ as a function of $S^z$ for $s=\frac{1}{2}$ and and $J'$ equal to (a) $\frac{J}{50}$, and (b) $\frac{3J}{5}$. The spin indices of the legend (and the color coding) refer to Fig. \ref{fig:twododecahedraquantum}.
%(/basic/diag/dodectwo)
}
\label{fig:spinzquantum}
\end{figure}

Fig. \ref{fig:spinzquantum} shows the distinct ground state expectation values of the projections of the individual spins along the $z$ axis $<s_i^z>$ for $J'=\frac{J}{50}$ and $\frac{3J}{5}$. There are in principle four unique such projections, with spins 1 to 5 having a common one (Fig. \ref{fig:twododecahedraquantum}), and spins 16 to 20 another one. Also every second of spins 6 to 15 shares the same value of $<s_i^z>$. Spins of the central and the outer pentagon have lower values for $J'=\frac{J}{50}$, which agrees with their stronger intradodecahedral AFM correlations of Fig. \ref{fig:correlationsquantum}(a). For $J'=\frac{3J}{5}$ the central pentagon has even lower $<s_i^z>$, which now corresponds to the strongest AFM correlations being between spins in this pentagon, and between spins in this pentagon and their counterparts in the corresponding pentagon of the other dodecahedron connected via the $J'$ bonds, as shown in Fig. \ref{fig:correlationsquantum}(b).

\section{Conclusions}
\label{sec:conclusions}

The ground state magnetic response of two coupled dodecahedra was investigated within the framework of the AHM. The classical magnetization discontinuities of an isolated dodecahedron were found to be renormalized by the interdodecahedral coupling, while new ones emerge. For a specific range of the coupling the total number of discontinuities goes up to six. At the full quantum limit $s=\frac{1}{2}$ the isolated dodecahedron magnetization discontinuity gives rise to two neighboring discontinuities, with a third one appearing adjacent to these two. The two dodecahedra system has a magnetic response which for a significant range of the field is associated with magnetization steps with $\Delta S^z=2$, which is twice as strong as the usual magnetization difference between adjacent $S^z$ sectors for a quantum spin system. This shows that the magnetization change can be controlled by adjusting the range of an external magnetic field.

The frustrated nature of the dodecahedron results in unexpected ground state magnetization discontinuities in a field when the AHM is considered on it. Usually such discontinuities are associated with magnetic anisotropy, but in this case they are allowed by the special connectivity of the dodecahedron. The formation of a two dodecahedra molecule with the introduction of unfrustrated coupling between one of their faces enriches the ground state magnetic response. Along these lines, it is of interest to extend this investigation on more than two dodecahedra linked together to form a chain-type or even more complicated structures, and calculate the magnetic response while the individual cluster frustration and the coupling between clusters compete in the presence of an external magnetic field.

\bibliography{dodecahedra}

\newpage

\begin{table}[h]
\begin{center}
\caption{First order perturbation theory energy correction $E_1$ for the different $S^z$ sectors for the two coupled dodecahedra. The numbers were generated with double precision but are presented with five significants digits for the sake of brevity where applicable.}
\begin{tabular}{c|c|c|c|c|c}
$S^z$ & $E_1$ & $S^z$ & $E_1$ &
$S^z$ & $E_1$ \\
\hline
0 & 0 & 7 & 0.0045106 & 14 & 0.51917 \\
\hline
1 & -0.15988 & 8 & $\frac{1}{5}$ & 15 & 0.47352 \\
\hline
2 & 0.0052328 & 9 & 0.11821 & 16 & 0.62767 \\
\hline
3 & -0.064727 & 10 & $\frac{3}{10}$ & 17 & 0.49119 \\
\hline
4 & -0.083283 & 11 & 0.32070 & 18 & 0.88820 \\
\hline
5 & -0.16495 & 12 & $\frac{9}{20}$ & 19 & 0.88820 \\
\hline
6 & -0.0039974 & 13 & 0.48578 & 20 & $\frac{5}{4}$ \\
\end{tabular}
\label{table:firstordercorrection}
\end{center}
\end{table}

%\begin{figure}
%\includegraphics[width=3.5in,height=2.6in]{paperseven-1new}
%\caption{Configuration of phase I. The direction of each spin unit vector is
%  given as a pair of its polar and azimuthal angle, $(\theta_i,\phi_i)$.}
%\label{fig:001}
%\end{figure}

\end{document}